# Observation and resonant x-ray optical interpretation of multi-atom resonant photoemission effects in O 1s emission from NiO

N. Mannella[1,2,#], S.-H. Yang[2,†], B. S. Mun[1,2,4], F. J. Garcia de Abajo[2,3], A. W. Kay[1,2,‡], B.C. Sell[1,2], M. Watanabe[2,5], H. Ohldag[4,6] E. Arenholz[4], A.T. Young[4], Z. Hussain[4], M. A. Van Hove[1,2,§], and C. S. Fadley[1,2]

[1]*Department of Physics, University of California-Davis, Davis, California 95616*
[2]*Materials Sciences Division, Lawrence Berkeley National Laboratory, Berkeley, California 94720*
[3]*Centro Mixto CSIC-UPV/EHU, San Sebastian, Spain*
[4]*Advanced Light Source, Lawrence Berkeley National Laboratory, Berkeley, California 94720*
[5]*RIKEN, Hyogo 679-5148, Japan*
[6] *Stanford Synchrotron Radiation Laboratory, P.O. Box 20450 Stanford, California 9430*
[#] *Present address: Physics Department, Stanford University, Stanford, CA, USA*
[†] *Present address: IBM Almaden Research Center, San Jose, CA 95120, USA*
[‡]*Present address: Intel, Hillsboro, OR 89712 USA*
[§]*Present address: Dept. of Physics and Materials Science, City University of Hong Kong, Kowloon, Hong Kong, China*

## ABST RACT

We present experimental and theoretical results for the variation of the O 1s intensity from a NiO(001) surface as the excitation energy is varied through the Ni $2p_{1/2,3/2}$ absorption resonances, and as the incidence angle of the radiation is varied from grazing to larger values. For grazing incidence, a strong multi-atom resonant photoemission (MARPE) effect is seen on the O 1s intensity as the Ni 2p resonances are crossed, but its magnitude decreases rapidly as the incidence angle is increased. Resonant x-ray optical (RXRO) calculations are found to predict these effects very well, although the experimental effects are found to decrease at higher incidence angles faster than those in theory. The potential influence of photoelectron diffraction effects on such measurements are also considered, including experimental data with azimuthal-angle variation and corresponding multiple-scattering-diffraction calculations, but we conclude that they do not vary beyond what is expected on the basis of the change in photoelectron kinetic energy. Varying from linear polarization to circular polarization is found to enhance these effects in NiO considerably, although the reasons are not clear. We also discuss the relationship of these measurements to other related interatomic resonance experiments and theoretical developments, and make some suggestions for future studies in this area.

# I. INTRODUCTION

Resonant photoemission (RPE) is a very important spectroscopic tool for investigations of electronic interactions in matter. In the most common form of such measurements, the photon energy is tuned through a core-level absorption edge in a certain atom, and the photoemission cross section for another less-tightly-bound core or valence level in the same atom is observed to be significantly affected, with both increases and decreases being seen, and the energy dependence of the effect usually following the well-known Fano profile[1]. Among other things, RPE can thus be used to distinguish the contributions of individual components in the valence bands.

More recently[2,3,4,5,6,7], it has been pointed out that photoemission associated with a certain core electronic level of a given atom "A" can be significantly altered in intensity by tuning the photon energy through core-level absorption edges of a near-neighbor atom "B". This effect has been termed *multi-atom resonant photoemission* (MARPE) to distinguish it from the *single-atom* aspect of normal RPE (SARPE). Although some first measurements and discussions of these effects were influenced by the non-linearity of a particular detector system utilized[2,3,5,6,8,9], it has subsequently been found that the experimental MARPE effects observed for photoemission from a homogeneous semi-infinite solid surface (in particular MnO(001)) can be predicted using either a microscopic interatomic resonant photoemission theoretical model or a macroscopic x-ray optical (dielectric) model whose input is the complex dielectric constant on passing the relevant resonance[4,7,10]. We will refer to the latter model as a *resonant x-ray optical* (RXRO) approach. The former microscopic model has also been shown, for the specific case of semi-infinite layer systems and with the inclusion of higher-order x-ray interactions not considered previously, to be reducible to the x-ray optical treatment using experimental optical constants[4,7]. The RXRO model is furthermore found to well describe the observed intensity profiles as a function of both photon energy and x-ray incidence angle. More specifically, both the microscopic resonant photoemission model and the RXRO model are found to well reproduce experimental data observed for the case of O $1s$ emission from MnO[7] on passing the Mn $2p_{1/2,3/2}$ resonances, with the microscopic model also having been shown to explain analogous data for O $1s$ emission from CuO on passing the Cu $2p_{1/2,3/2}$ resonances[11]. In a more recent study of MARPE effects for $N_2$ adsorbed on Ni(111), it has been found that N $1s$ emission on passing the Ni $2p_{1/2,3/2}$ resonances also can be described in terms of the RXRO model[12]; although this study also attempted to observe short-range MARPE effects that went beyond the RXRO picture, these were not found, at least to within ~1% in magnitude.

As noted previously, it has also been found in such studies to be very important to allow for any detector non-linearity present, in measuring either photoelectron intensities or x-ray absorption coefficients via secondary electrons, since the flux of both primary and secondary electrons increases dramatically over any strong absorption edge, thus forcing the detection system to span a large dynamic range[7]. Accurate procedures have been developed for such corrections and demonstrated experimentally[7,13,14,15].

Related studies on solids have also dealt with the case of soft x-ray emission under conditions of interatomic resonance, specifically from MnO[6], $LaF_3$[16] and $Ti_xNb_{1-x}C$[16]. For these cases, the RXRO model was found to describe the data well, although the connection of this model with interatomic resonant photoemission was not fully recognized in some of this work[16].

MARPE effects have also been considered for small molecules, from both theoretical and experimental points of view. In such cases, the effect must arise strictly from short-range interactions over the few neighboring atoms available, rather than being summed over many atoms as in an extended solid, as discussed in prior work[4]. An alternative microscopic theoretical formulation that is in fact closely related to that presented in ref. 4 has been discussed[17], and this work concludes that the effects on experimental intensities may be as small as 1%. However, this study underestimates the relative intensity of such effects in considering the <u>squares</u> of the direct and resonant matrix elements involved, rather than the matrix elements themselves, since it is the interference between the two processes that leads to MARPE[4,10].

Finally, more subtle effects on the *non-dipole* angular distributions parameters in core emission from small gas-phase molecules have been observed, in particular for N $1s$ emission from $N_2O$ with two inequivalent N atoms as the O $1s$ resonance is scanned, and C $1s$ and S $2p$ emission from OCS, again as the O $1s$ resonance is scanned[18]. These effects have been termed near-atom core-hole transfer (NACHT), but they also represent a manifestation of MARPE, albeit a more subtle one.

We here treat the case of NiO, which is of interest since a previous investigation found that the O$1s$ photoemission intensity from NiO(100) did not show *any* modulation on passing the Ni $2p_{1/2,3/2}$ resonances[11]. This lack of any effect was initially attributed to some fundamental difference between NiO and MnO or CuO, but it has subsequently been suggested that it is due simply to the relatively high x-ray incidence angle (35°) used in these measurements, for which RXRO theory predicts a falloff in the fractional changes in intensity[7]. We present new experimental evidence to support this suggestion, compare the data with RXRO theory, and discuss other implications and applications of such phenomena, including the influence on such effects of photoelectron diffraction (PD) and variable polarization.

## II. EXPERIMENTAL

The experiments were carried out using the Multi-Technique Spectrometer/Diffractometer[19] (MTSD) system located on beamline 4.0.2 at the Berkeley Advanced Light Source[20], where photoelectron spectra were measured with a Scienta SES200 electron spectrometer. In previous work[7,8(b),13,14,15], it has been pointed out that a proper allowance for detector non-linearity is essential for accurately measuring intensities with the standard detector system supplied with this spectrometer. We have for the present study calibrated our detector system in the "analogue" ("greyscale") mode by using a standard x-ray tube with continuously-variable emission current at fixed high voltage. Details concerning this calibration and its resulting correction procedure for non-linearity effects appear elsewhere[7,13,14,15].

A NiO (001) surface was prepared by initially cleaving a single crystal in air and then immediately transferring it via a UHV-compatible loadlock into the sample preparation chamber of the Multi-Technique Spectrometer / Diffractometer (MTSD) system. Subsequent *in situ* cleaning consisted of ion sputtering (Ar, 15 minutes at 500 V beam voltage) and then annealing in oxygen at $1\times10^{-6}$ Torr and 923 K for 3 hours. An x-ray photoelectron spectroscopy (XPS) analysis confirmed the absence of surface contamination and the expected stoichiometric ratio between Ni and O. Low energy electron diffraction (LEED) and photoelectron diffraction (PD) patterns obtained immediately after the treatment showed all the features expected from the (001) surface of NiO, which has a rock-salt structure.

The sample was kept at a constant elevated temperature of 653 K in order to avoid charging effects due to the insulating nature of NiO at room temperature, a temperature which is also above the antiferromagnetic transition temperature of $T_N = 524$ K[21]. In a first set of data, the photoemission spectra were collected with the take-off angle of the photoelectrons set to 90° with respect to the surface of the sample (i.e. normal emission), thus yielding the maximum average electron escape depth. For most measurements, the sample was oriented azimuthally such that the plane containing the x-ray incidence direction and the photoelectron emission direction contained the [100] direction in the sample surface, as shown in Fig. 1(a). Both x-ray incidence angle $\theta_{hv}$ and photon energy were varied in these measurements, with particular emphasis on the absorption edges in the Ni $2p_{1/2,3/2}$ region in energy. In a second set of data, the x-ray incidence angle was set at 20º and the photoelectron take-off angle $\theta_e$ at 45º and the sample rotated so that the photoelectron azimuthal angle $\phi_e$ was varied over 360º. In this second set, the modulations of the O 1s intensity due to photoelectron diffraction were measured at different photon energies, and for different incident polarizations. For reference, the Fermi-referenced O 1s binding energy we measure to be 532 eV, in agreement with prior work[11].

## III. RESULTS AND DISCUSSION

### IIIA. The Detection of MARPE effects in NiO

In Fig 1(b), we show O1s intensities, fully corrected for detector non-linearity effects and measured as areas by fitting Voigt peak shapes with Shirley inelastic backgrounds to O 1s photoemission spectra, as a function of photon energy and for different values of the x-ray incidence angle $\theta_{hv}$ between the direction of the incoming beam and the surface of the sample. The photon energy has here been scanned in small steps of 0.25eV over the region of the Ni $2p_{3/2}$ and Ni $2p_{1/2}$ absorption peaks. These curves show the same qualitative signatures seen in analogous previous measurements of O 1s emission from MnO[7] on passing the Mn 2p edges and CuO on passing the Cu $2p_{3/2}$ edge[11]. The intensity shows a decrease just below each of the two absorption resonances and then an increase on going above it. Moreover, in agreement with similar data obtained for MnO over x-ray incidence angles over the range of 5º-30º, the effects are strongly dependent on angle, being largest for more grazing x-ray incidence angles, and quickly decaying in magnitude as this angle is increased[7]. The overall effect, as judged by [(intensity maximum just above Ni $2p_{3/2}$) - (intensity minimum just below Ni $2p_{3/2}$)]/(average intensity for a smooth curve passing through the Ni $2p_{3/2}$ resonance) is quite similar to MnO, ranging from about 5% for the highest incidence angle of 40° to about 70% for the lowest incidence angle of 5°. Without MARPE (or RXRO) effects, one should observe a simple smooth curve of negative slope over this region in energy due to a combination of subshell cross section[22] and electron inelastic mean free path (IMFP) variations[23], as perhaps modulated by energy-dependent photoelectron diffraction (PD)[24]. Although the effective attenuation length (EAL) may differ somewhat from the IMFP due to effects of elastic scattering, for our conditions of relatively low-atomic number scatterers, electric field vector within 40º of the electron emission direction, and emission direction along the normal, we can assume that IMFPs and EALs are fairly close to one another[23(b)].

These data by themselves thus demonstrate that earlier measurements on NiO[11] somehow missed the presence of MARPE (or RXRO) effects, probably due to being carried out at higher incidence angles and with insufficient statistical accuracy to see the small effects present. We now turn to the theoretical interpretation of these data based on the RXRO model.

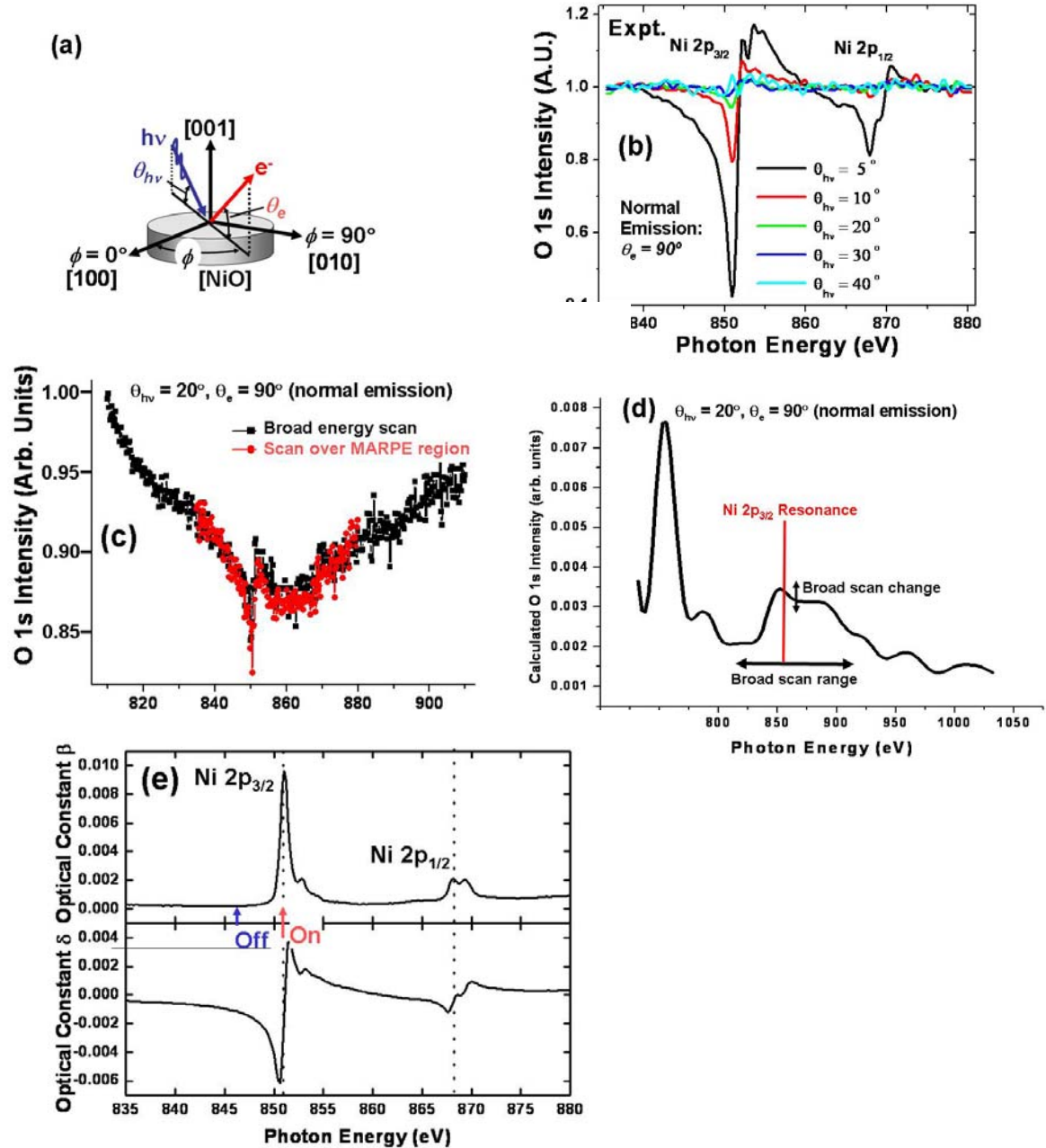

**Figure 1 (Color online): (a)** Experimental geometry. **(b)** Summary of experimental data for O 1$s$ intensity as a function of photon energy on passing through the Ni 2$p$ absorption-edge resonances, for several angles of x-ray incidence and photoelectron

emission along the NiO(100) surface normal. **(c)** Comparison of the experimental data from (b) for an incidence angle of 20° with similar data over a more extended energy range, with the curvature being due to scanned-energy photoelectron diffraction effects. **(d)** Multiple scattering photoelectron diffraction calculations of the energy dependence of O 1s intensity from NiO(001), carried out over a very broad energy range encompassing the broad scan in (c). The energy range and fractional excursion of the data in (c) are also indicated. **(e)** The optical constants $\beta$ (determined from partial-electron-yield x-ray absorption data) and $\delta$ (determined via Kramers-Krönig analysis of $\beta$) as a function of photon energy in crossing the Ni 2p resonances.

### IIIB. The RXRO Theoretical Model

As mentioned above, a microscopic quantum-mechanical model of MARPE previously developed in our group for describing similar experimental results[4,7] in other systems has been found to well describe such effects, for example, for the specific case of O 1s emission from MnO in the vicinity of the Mn 2p resonances, confirming via agreement with experiment that these effects can be considered as interatomic resonance photoemission phenomena. This microscopic model was also found to be reducible to a RXRO treatment using experimental optical constants. We refer the reader to previous work for details on both of these models and the discussion of their equivalence under certain conditions[4,7], but focus for interpreting the present data on the RXRO approach, making use of a general-purpose program written by Yang[7(b)].

In the RXRO approach, the effect of the resonance is assumed to influence only the local electric field $\vec{E}$ at some depth $z$ below the homogeneous flat surface of the sample, with the differential photoelectric cross section $d\sigma/d\Omega$ varying only slowly through the resonance as described by the usual one-electron theory[7]. The variation of photoemission intensity with photon energy $I(h\nu)$ is then obtained by integrating over the coordinate $z$ perpendicular to the surface (here taken to increase into the surface) the product of the electric field strength $|\vec{E}(h\nu,z)|^2$ at depth $z$ relevant for photoemission, the energy-dependent differential photoelectron cross section $d\sigma/d\Omega$ appropriate to the experimental geometry (which may also in the experimental data include the effects of photoelectron diffraction) and the kinetic-energy-dependent IMFP for electrons $\Lambda_e$, as

$$I(h\nu) \propto \frac{d\sigma(\hat{E},h\nu)}{d\Omega} \int_0^\infty |\vec{E}(h\nu,z)|^2 \exp(-\frac{z}{\Lambda_e(E_{kin})\sin\theta_e})dz \qquad (1)$$

where $\hat{E}$ is a unit vector along $\vec{E}$ and accounts for the polarization dependence in the cross section, and we have not included factors of atomic density and the solid angle acceptance of the analyzer that will be constant over an energy scan.

The effect of scanned-energy PD[24] on the O 1s intensity mentioned above appears to be evident in Fig. 1(c), where a narrow energy scan like that in Fig. 1(b) and also for an incidence angle of 20° is compared to a much wider energy scan. There is a clear slowly varying modulation of the intensity by about 18% in the wider scan, similar to that observed previously in MnO (cf. Fig. 1(e) of Ref. 7). To further illustrate the potential influence of PD on such energy scans, we show in Fig. 1(d) multiple-scattering PD calculations for a cluster of 250 atoms representing a NiO(100) surface, making use of the recently developed EDAC program[25]. The modulations predicted here are substantial, and we have indicated the energy range of the broad scan in Fig. 1(c), as well as its fractional change in Fig. 1(d). Although the exact form of the variation in Fig. 1(c) is not predicted by theory, it is clear that PD effects are easily large enough to explain this variation. The lack of better agreement is perhaps due to using too small a cluster. In any event, in subsequent comparisons of experiment and RXRO theory, we have thus divided out a smooth curve from the experimental data so as to focus more clearly on only the multi-atom resonant effects.

Calculating the electric field strength at depth $z$ is carried out via a knowledge of the photon-energy dependent index of refraction $n_r(h\nu)$, which is in turn related to the x-ray optical constants $\delta(h\nu)$ and $\beta(h\nu)$ as $n_r = \sqrt{\varepsilon} = 1 - \delta + i\beta$, with $\beta$ being determined by measuring the absorption coefficient $\mu(h\nu) = 4\pi\beta(h\nu)/\lambda_x$ over the edges in question (Ni 2p in our case). The absorption coefficient $\mu(h\nu)$, obtained by measuring inelastically scattered electrons of 175 eV kinetic energy, included an extrapolation of the absorption curves measured as a function of the x-ray angle of incidence to obtain the most accurate result[26]. Finally, $\delta$ was derived from $\beta$ using a Kramers-Krönig transformation. The x-ray optical constants $\delta(h\nu)$ and $\beta(h\nu)$ as derived experimentally in this study are shown as a function of photon energy in Fig. 1(e). Note also that the variation in the experimental O 1s intensity in Fig. 1(b) about a mean value follows very closely the behavior of $\delta$, just as observed in previous experimental data on MnO[7], a point on which we comment below. From the values of $\delta$ given in Fig. 1(e), we can furthermore estimate that the maximum critical angle for the onset of total reflection will be at the Ni 2p$_{3/2}$ resonance and will be given by the usual formula as $\theta^c_{inc} = \sqrt{2\delta} \approx 4.8°$.

Via an analysis based on the Fresnel equations[7,27], it can finally be shown that the integral in Eq. (1) reduces to

$$I(h\nu) \propto \frac{d\sigma}{d\Omega}(\hat{E}, h\nu) \frac{|t(h\nu)|^2}{\frac{\text{Im}\{4\pi n_r(h\nu)\sin\theta'_{h\nu}(h\nu)\}}{\lambda_x(h\nu)} + \frac{1}{\Lambda_e(E_{kin})\sin\theta_e}} \quad (2)$$

where the quantity $t$ for p-polarized radiation incident on a planar surface from vacuum with $n = 1$, and for a conducting or non-conducting, but non-magnetic, reflective medium, is given by

$$t = \frac{2\sin\theta_{h\nu}}{\sin\theta'_{h\nu} + n_r\sin\theta_{h\nu}} \quad (3)$$

with $\theta_{h\nu}'$ equal to the complex angle of propagation below the surface, again measured relative to the surface, and $\lambda_x$ the wavelength of the radiation. $\theta_{h\nu}'$ is further related to $\theta_{h\nu}$ via Snell's Law: $\cos\theta_{h\nu} = n_r\cos\theta'_{h\nu}$, with $\theta_{h\nu}$ real. Eqs. 2 and 3 are thus very general formulas for calculating photoemission intensity from a semi-infinite substrate, with all dependences on energy explicitly indicated. Beyond the optical constants of Fig. 1(e), the only other inputs needed are radial matrix elements and phase shifts for calculating $d\sigma/d\Omega$[22] and the electron IMFP $\Lambda_e$, which we have evaluated for the O 1s photoelectrons leaving NiO using the well-established TPP-2M semi-empirical formula[23]. $\Lambda_e$ is found to vary only slightly, from 8.0 Å to 8.7 Å, over the energy region covered in our measurements.

## IV. DISCUSSION OF RESULTS
### IVA. Effects at Fixed Emission Angle

Before proceeding to compare the results of Fig. 1(b) with RXRO theory, we present some basic theoretical results to illustrate some of the physical effects arising. In Fig. 2(a), we show the exponential decay length of the incident radiation perpendicular to the surface as a function of incidence angle, for two photon energies, one below and the other on the Ni 2p$_{3/2}$ absorption edge. It is clear that the penetration depth is reduced markedly on going to very low angles of incidence and that is it much lower on the resonance, in fact approaching the electron IMFP ($\approx 8$ - 9 Å as estimated from the TPP-2M formula[23]) in magnitude. This will thus significantly enhance the surface sensitivity of the measurement, and it is one effect that will automatically be included in the model of equations (2) and (3). Next in Fig. 2(b), we show the variation of reflectivity as a function of both incidence angle and photon energy, with this plot making it clear that reflectivity is significantly enhanced both by going to lower incidence angles (an expected effect) and by crossing a strong absorption resonance.

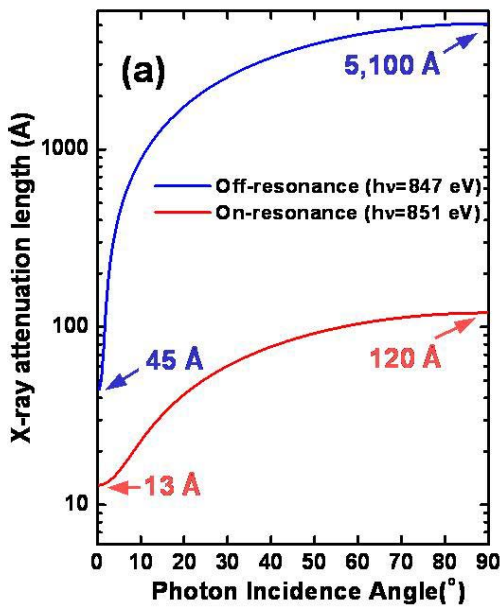
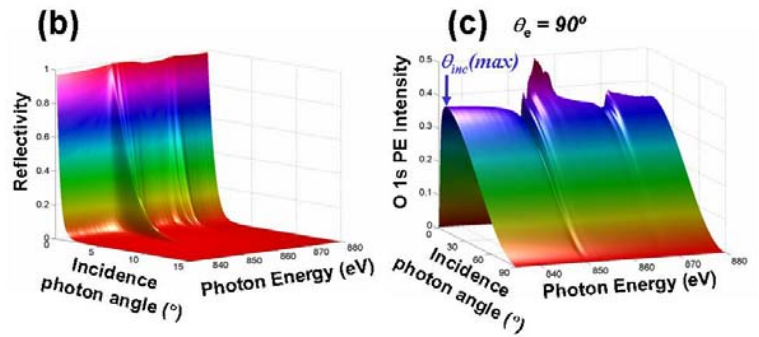

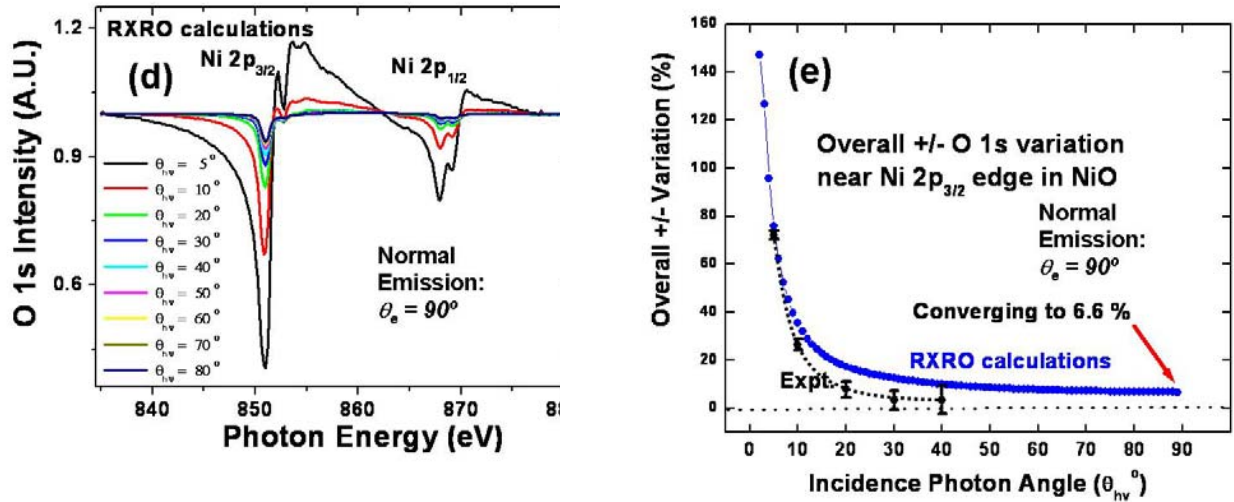

**Figure 2 (Color online): (a)** Theoretical resonant x-ray optical (RXRO) calculation of the exponential decay length normal to the surface of the square of the electric field, for photon energies off resonance (847 eV) and on resonance (851 eV). **(b)** RXRO calculation of reflectivity as a function of both photon energy and incidence angle as the photon energy crosses the Ni $2p_{1/2,3/2}$ absorption resonances. **(c)** As (b), but for O 1s intensity, with the incidence angle at which the maximum intensity is found indicated as $\theta_{inch}(max)$. **(d)** RXRO calculations of the effects seen in Fig. 1(b). **(e)** Comparison of theoretical and experimental results for the overall amplitude of the plus/minus variation of the effect in crossing the Ni $2p_{3/2}$ resonance as a function of incidence angle, expressed as a percentage of the intensity at 830 eV below the resonance

Next, we consider RXRO calculations of the actual O 1s photoelectron intensity, as summarized over the full range of incidence angles and photon energies in Fig. 2(c), as well as for the specific cases for which we have experimental data in Fig. 2(d). The overall amplitude of the plus/minus variation of the effect in crossing Ni $2p_{3/2}$ as a function of incidence angle is further shown in Fig. 2(e), where it is expressed as a percentage of the intensity at 830 eV below the resonance, and compared to experimental results at the five incidence angles studied. The results in these three figures make it clear that the resonant effects are strongly sensitive to x-ray incidence angle, being much smaller for angles greater than about 30 degrees, although very similar in energy dependence for all angles. Going to an incidence angle of 5° is also predicted to yield a very large effect of ~70 %, and this is observed in our experiments. Small effects of about 5% are also predicted to persist up to normal emission for the Ni $2p_{3/2}$ case. Comparing the experimental results in Fig. 1(b) with the theoretical curves in Fig. 2(d) further shows excellent agreement as to the qualitative behavior. Finally, the absolute O 1s intensity in Fig. 2(c) is found to have a maximum value at an incidence angle of about 11°.

As a more quantitative comparison of experiment and RXRO theory, Figs. 3(a)-3(e) directly compare the experimental data from Fig. 1(b) with corresponding RXRO theory. The experimental data have been divided by a smooth curve to reduce PD effects, while the theoretical curves have been divided by a linear slope in order that both experiment and theory agree below and above the resonances. Finally, all curves were normalized to unity at the far left of the scan, below the resonances. There is excellent agreement between experiment and theory for the two lowest incidence angles of 5° and 10°, and semiquantitative agreement for the higher angles of 20°, 30°, and 40°, with the agreement deteriorating as the angle increases. Theory generally predicts larger effects than those seen in experiment for incidence angles above 5° and furthermore yields an asymptotic value at normal incidence of about 6.6 % for the overall effect that seems not to be consistent with the experimental trend up to a 40° incidence angle. In particular, the observed effects are only about 1/3-1/2 as large as predicted by RXRO theory for the highest three incidence angles. Since the surface under study is very flat (visually mirror-like and yielding sharp

LEED patterns) and stoichiometric, with no traces of contaminants (as confirmed by quantitative XPS analysis), this type of disagreement could be due to the lack of using the more correct microscopic theory of MARPE, a point to which we return below. As a qualitative indicator of what the results at higher incidence angles might imply about the effective dielectric constants involved, Fig. 3(f) compares the experimental MARPE results for an incidence angle of 20º with RXRO calculations in which both δ and β appearing in the experimental dielectric constant have simply been scaled down by a variable factor. For this angle of incidence, scaling down by about 50% is found to yield reasonable, though not perfect, agreement with experiment. This comparison suggests that microscopic and/or local-field effects in the near-surface region (which deviate from the macroscopic approach represented by the RXRO model) are responsible, but further experimental and theoretical study is necessary to understand these deviations between experiment and RXRO theory.

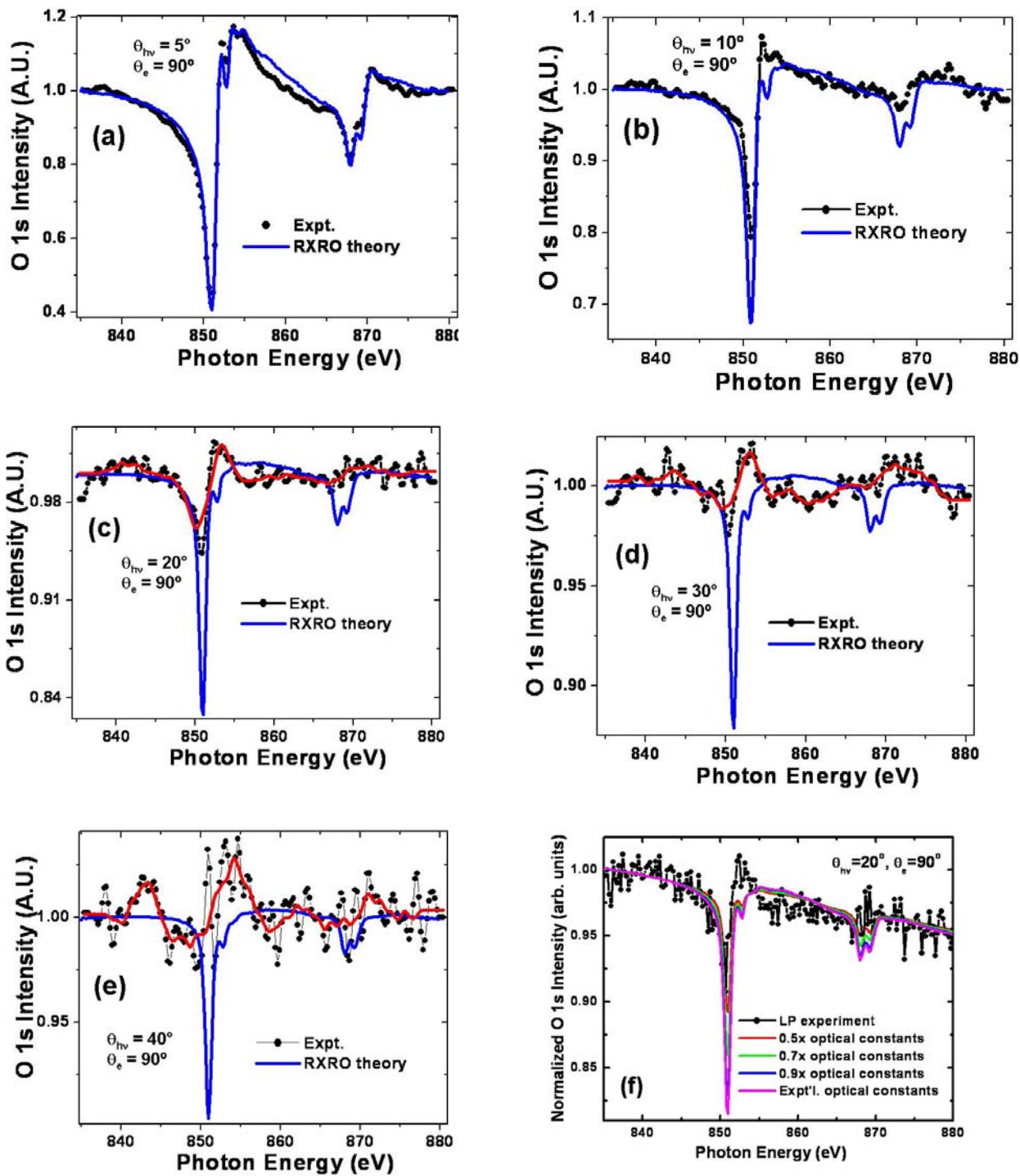

**Figure 3 (Color online):** **(a)-(e)** Direct comparison of experimental O 1*s* intensities on passing the Ni 2*p* resonances with RXRO theory, with experiment having been corrected for curvature due to photoelectron diffraction (cf. Figs. 1(c) and 1(d)) and theory and experiment having been brought into agreement below and above these resonances by a suitable division by a linear function. Otherwise, no scaling has been done, so that percentage changes are correct. In Figs. (c)-(e), red curves representing ten-point (11[th]-order) Savitzky-Golay smoothing of the data are included to more clearly show the small effects seen, due to the greater statistical scatter of the data. **(f)** Comparison of expt. with RXRO theory for an incidence angle of 20º. Here, the theoretical curves are shown for various multiplicative reductions of the optical constants in Fig. 1(e), as explained in the text

In general, however, these results thus further confirm the general validity of the x-ray optical analysis in providing a phenomenological and semi-quantitative description of the experimental MARPE effects, and also imply that such effects should be observable on crossing strong core-level resonances for <u>all</u> angles of x-ray incidence, although with greater difficulty of observation as $\theta_{h\nu}$ goes above about 20°. These comparisons of experiment and theory thus further confirm the conclusion that no MARPE effect was seen in NiO in the prior study mentioned[11] above because of the high incidence angle used. We also note that a prior unsuccessful attempt to see MARPE effects in magnetic thin films also was no doubt also influenced by the high angle of incidence of 45º that was used[28].

A further suggestion based on this work is that such experimental data could be used as another means of determining $\delta(h\nu)$ over such resonances, for example, by measuring $\beta(h\nu)$ and then choosing $\delta(h\nu)$ so as to yield the best fit to $I(h\nu)$ for incidence angles less that about 10°. However, our data also suggest that a fully accurate prediction of these effects may require a more detailed microscopic theory of the local fields near a surface, such as that discussed previously[4,7] or in fact going beyond that description, with other approaches including a detailed discussion of electron screening in resonant photoemission and a full relativistic treatment of the electronic states having been presented recently[10]. An additional uncertainty in theory will always be associated with the electron inelastic attenuation length.

As one further point, we comment on a couple of approximate limits of Eq. (2) that help to explain the qualitative form of the curves in Figs. 1(b), 2(d), and 3. In one limit, far away from the total reflection region ($\theta_{h\nu} \geq 70°$), several simplifications yield finally[4,7]

$$I(h\nu) \approx \frac{d\sigma(\hat{E},h\nu)}{d\Omega} \cdot \frac{1+\delta(h\nu)}{\frac{4\pi\beta(h\nu)\sin\theta_{h\nu}}{\lambda_x(h\nu)} + \frac{1}{\Lambda_e(E_{kin})\sin\theta}} \quad . \quad (4)$$

The first term in the denominator is smaller than the second by at least a factor of 17, with its maximum value relative to the second term being at the maximum of β (~0.01 as seen in Fig. 1(e)) and for normal incidence and electron takeoff. Also, the second term in the denominator is essentially constant over the small energy range scanned in our data. Thus the variation of β in the denominator can give rise to about a 6-7 % modulation of the photoelectron intensity, which is very close to that predicted at higher takeoff angles in Fig. 2(e) and 3(e). The magnitude of this variation is also enhanced slightly by the change in δ due to the numerator *1 + δ(hν)*, but this makes a contribution of only about 1% to the variation of intensity (again see Fig. 1(e)). This result explains why the intensity variation for higher incidence angles is qualitatively similar to that of an inverted curve of β.

Looking now at another limit of Eq. 2 that is valid as one <u>approaches</u> the total-reflection regime, corresponding to $2\delta \leq \theta_{h\nu}^2 \ll 1$, one finds after some manipulation an expression involving three factors:

$$I(h\nu) \approx \frac{d\sigma(\hat{E},h\nu)}{d\Omega} \cdot \frac{4}{\left(1+\sqrt{1-\frac{2\delta}{\theta_{h\nu}^2}}\right)^2} \cdot \frac{1}{\frac{4\pi\beta(h\nu)\sin\theta_{h\nu}}{\lambda_x(h\nu)} + \frac{1}{\Lambda_e(E_{kin})\sin\theta_e}} \quad . \quad (5)$$

As the incidence angle and thus also $sin\theta_{h\nu}$ decreases, the denominator in the third factor is dominated by the essentially constant second term $1/(\Lambda_e(E_{kin})sin\theta_e)$, while the denominator in the second factor starts making a greater difference. Consequently, the shape of the intensity begins to be modified from that of the inverse of the *β* curve, with the variation of *I(hν)* then mostly being determined by the second factor that contains only *δ*. If furthermore $\theta_{hv}^2$ is very close to $2|\delta|_{max}$ and $|\delta_{max}| \approx |\delta_{min}|$ (the values are actually +0.0035 and -0.006 in Fig. 1(e)), then the second factor may vary from $4/(1+\sqrt{2}))^2 = 0.68$ to 4 at maximum, yielding an overall estimated variation of 83% at the critical angle that is at least qualitatively consistent with Fig. 2(e). The variation in the experimental O 1s intensities as a function of the photon energy is thus qualitatively explained by Eq. (4) for high incidence angles and Eq. (5) for grazing incidence angles (but not in the total reflection region). It is not straightforward to obtain a simplified form of Eq. (2) in the intermediate regime or the total reflection regime.

## IVB. Resonant Effects on Photoelectron Diffraction

As another aspect of our data, we show in Fig. 4(a) azimuthal scans of O 1s intensity for a fixed x-ray incidence angle of 20° (for which the overall resonance effect in normal emission is about 8%) and a fixed photoelectron take-off angle of 45°. These data have been obtained at two photon energies, one below the Ni 2p resonances at 835 eV and one directly on the Ni $2p_{3/2}$ resonance at 851 eV. The solid-angle of acceptance of our spectrometer is approximately over a cone of 5° half-angle. The scanned-angle photoelectron diffraction effects themselves are dramatic, with modulations of over 50% relative to the maximum intensity and pronounced fine structure; this result attests to the high degree of order and cleanliness of this cleaved surface. Although the change in the O 1s photoelectron wavelength over this 16 eV energy change is not large (from 0.704 Å to 0.686 Å, or only about 3%), we still see significant changes in the fine structure located at the high-symmetry directions denoted by the arrows in Fig 4(a). Prior work on O 1s emission from MnO has also shown effects on photoelectron diffraction features along certain directions that are different along certain directions in going from below to on resonance that are possibly connected with MARPE[13]. Thus we ask whether these changes for the present data from NiO are due to interatomic resonance effects or are simply due to subtle differences in the photoelectron diffraction patterns at these two energies.

To assess the second possibility, we have again carried out multiple scattering photoelectron diffraction calculations with the EDAC program[25], this time for a cluster of 450 atoms representing a Ni(100) surface. Multiple scattering up to 25$^{th}$ order was included to insure convergence, and the calculations were averaged over the expected analyzer acceptance solid angle. Results for azimuthal scans at two different energies to simulate the data of Fig. 4(a) are shown in Fig. 4(b). It is first clear that these calculations do a very good job of simulating the observed fine structure and its changes on going from an off-resonance energy to an on-resonance energy. The doublet feature along the <100> azimuths is correctly predicted, as is the higher intensity of the features labeled with black arrows at the off-resonance energy. However, the changes in features with energy are somewhat exaggerated in theory, perhaps because the atomic cluster size was not large enough to adequately describe the diffraction patterns.

More detailed sets of experimental azimuthal scans and theoretical simulations for another NiO sample are presented in Figs. 4(c) and 4(d), respectively. The experimental data in Fig. 4(c), taken at emission angles from 41° to 47° in 2° intervals, make it clear that these photoelectron diffraction effects are very sensitive to emission angle. For example, the double-peaked structure along <100> is by far the most intense feature at 47°, but by 41° it is less intense than a single-peak feature that grows in 45° away in azimuth. Comparing experiment with multiple-scattering calculations for a 700-atom cluster that are shown in Fig. 4(d) again indicates that theory provides a good description of the trends with changing energy and emission angle, although theory again is found to exaggerate these trends since the calculations probably did not include a sufficient number of atoms in the cluster.

In summary, the general agreement found between experiment and theory on- and off- resonance does not permit concluding that there are any specific influences of interatomic resonant effects on photoelectron diffraction for NiO, although this warrants further study to be certain.

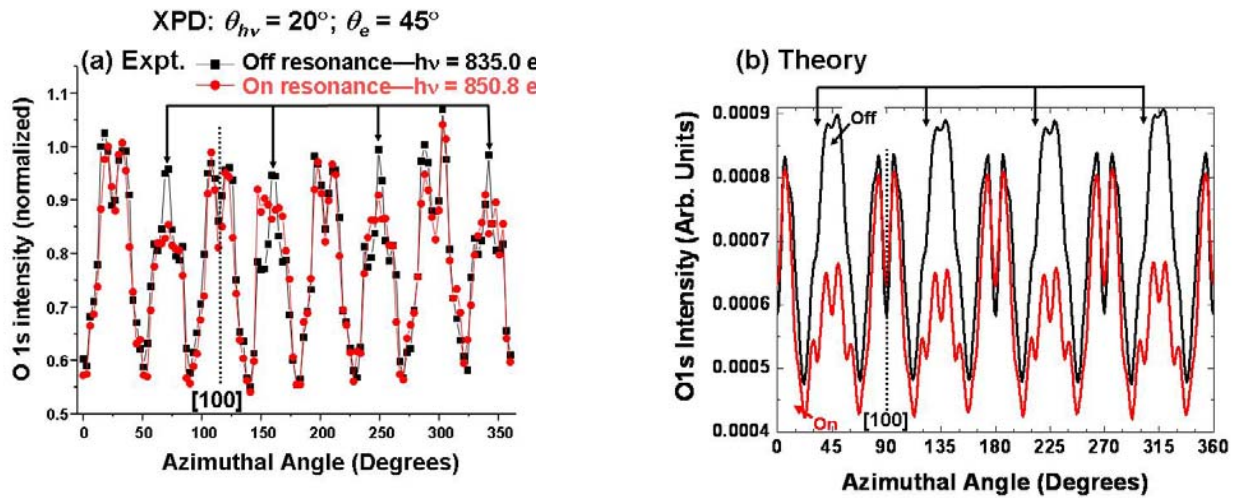

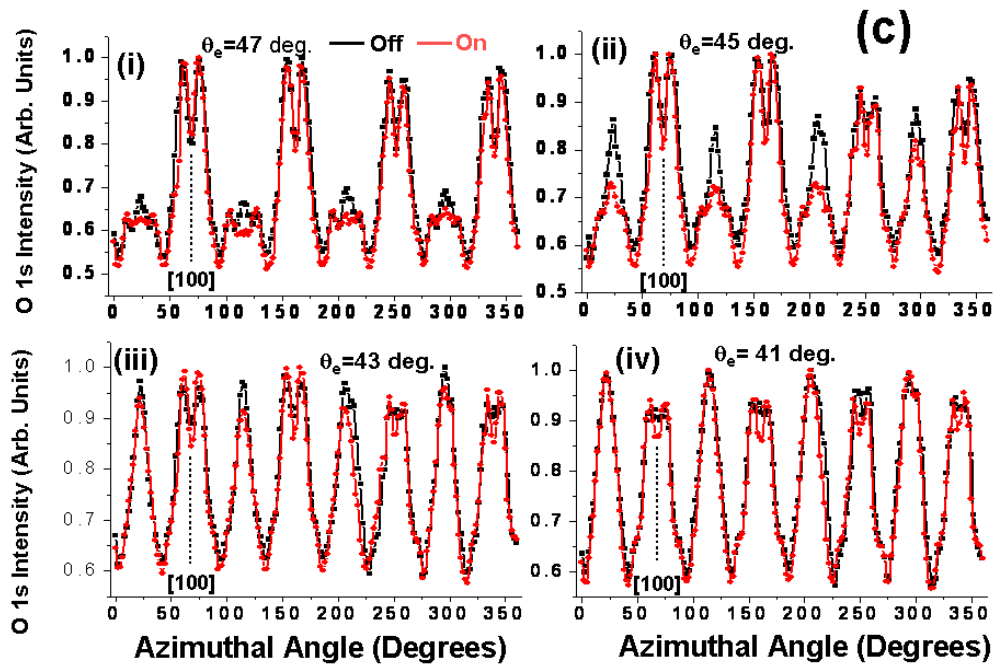
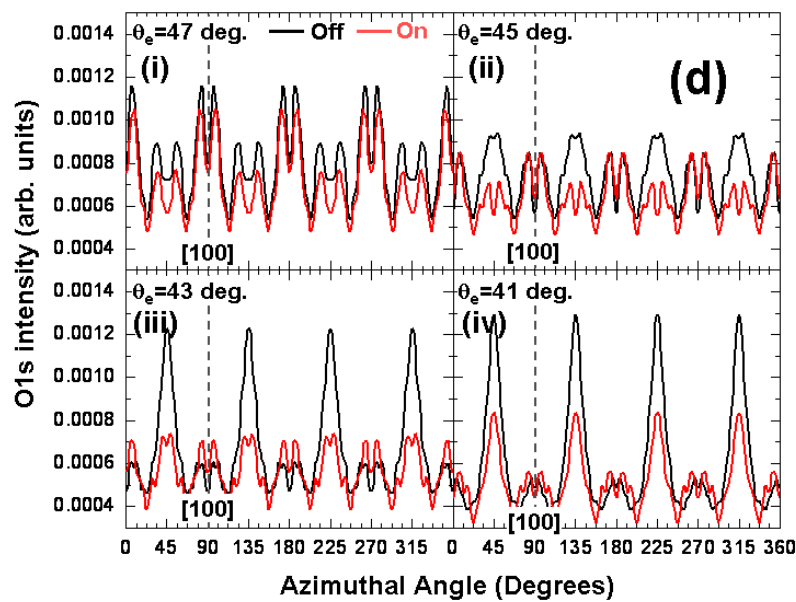

**Figure 4 (Color online): (a)** Scanned-angle O 1s photoelectron diffraction data for emission at a takeoff angle of 45° and two photon energies below (835.0 eV) and on (850.8 eV) the Ni $2p_{3/2}$ resonance labelled. A typical low-index <100> azimuth is labeled (cf. Fig. 1(a)), and four symmetry-identical features which change significantly on crossing the resonance are indicated by the arrows. **(b)** Multiple-scattering PD calculations of the curves in (a), plotted on the same horizontal scale. **(c)** Scanned-angle experimental data obtained from another sample at four equally-spaced takeoff angles from 41° to 47°. **(d)** Theoretical scanned-angle multiple-scattering photoelectron diffraction calculations over the same angle range as that in (c).

**IVC. Circular Dichroism Effects**

Finally, we consider the effect of using circularly polarized radiation for excitation in such MARPE scans, with results for the p-type linear-polarized (p-LP) case we have been discussing up to now being compared in Fig. 5(a) to analogous scans with right and left circular polarization (RCP and LCP). The incidence angle is 20° in all cases, and the electron emission direction is along the normal. Although all three of these curves are similar in form, there is a remarkable difference in the overall variation of them, with the previously-presented data p-LP showing an $\approx 7\%$ effect and the two curves for right- and left- circularly polarized light (RCP and LCP) showing nearly identical curves and much higher variations of about 23%. In addition, the CP data have distinctly different fine structure from the p-LP data.

The RXRO program we are using permits calculating the O 1s intensity for p-type LP, RCP and LCP light. However, if the optical constants shown in Fig. 1(e) (which were derived using secondary electrons excited by p-type LP light) are used for all three cases, the resultant curves for p-LP, RCP, and LCP are found to be identical, in disagreement with the marked differences seen in experiment. One obvious reason for this identity in theory is that the s-type LP that is added in with appropriate phase to make RCP and LCP light yields, via the dipole excitation process, a zero contribution to the O 1s cross section in the plane of observation. However, in the actual experiment, elastic scattering and diffraction can yield non-zero intensity along the surface normal via this component of polarization. An additional reason that experiment and theory as we have carried them out so far disagree is that the optical constants should be remeasured for CP radiation, and this we suggest for future experiments. However, since we are well above the Néel temperature of NiO, effects due to any sort of linear magnetic dichroism should be zero, and it is nonetheless not clear that one would expect much difference from the curves shown in Fig. 1(e).

Another obvious deficiency in these RXRO calculations is that they do not include the strong effects of photoelectron diffraction (cf. Figs. 4(a) and 4(c)). In fact, in combination with CP excitation, it is well known that circular dichroism in photoelectron angular distributions (CDAD) arises[29,30], with these effects also being describable in terms of forward scattering peak "rotations" in angle in the direction of rotation of the electric field vector[30]. Such rotation effects are obvious in the full-hemisphere multiple scattering calculations shown in Fig. 5(b), which have been carried out for a smaller cluster of 150 atoms, again both on-resonance and off-resonance in energy. However, since the dramatic difference is seen not in a dichroism signal, which would be very small if we subtract the two nearly identical curves for RCP and LCP in Fig. 5(a), but simply between excitation with LP and RCP $\cong$ LCP, it does not seem likely that photoelectron diffraction alone can explain the dramatic difference in the MARPE effects observed between linear- and circular-polarized excitations. In fact, the near identity of the RCP and LCP MARPE curves in Fig. 5(a) is indicative of a highly accurate alignment in azimuth along [100] (cf. Fig. 1(a)). These effects also are worth further experimental study.

In summary, varying polarization in such a core-level photoemission energy scan clearly can change the degree of MARPE effects dramatically, but it is still unclear as to what mix of not having the correct optical constants, not fully including photoelectron diffraction effects, and/or not having an accurate enough microscopic theory of the effects, is involved.

**V. CONCLUSIONS AND COMMENTS ON
OTHER RELATED MEASUREMENTS**

The experimental data for NiO presented here make it clear that multi-atom resonant photoemission (MARPE) effects are also observed in O 1*s* photoemission from this material, contrary to the conclusion reached in prior work[11]. These effects can be quantitatively explained, at least for x-ray incidence angles of $\leq \sim 20°$, via a resonant x-ray optical (RXRO) picture. Taken together with prior work by our group[7], these results lead us to conclude that similar effects should be seen for all materials, with the RXRO theory providing a more empirically-oriented method of analysis, and microscopic quantum-mechanical approaches outlined elsewhere[4,7,10] providing more generally applicable methods for treating not only homogeneous flat samples but nanostructures and free molecules. Such measurements also could provide an alternate method for determining the optical constants of materials on passing through core resonances, although with some caveats involving the details in the microscopic model (e.g. local-field and electron screening effects, the electron IMFP versus the EAL[23]). In making such measurements, the influence of photoelectron diffraction needs to be considered, although our analysis does not indicate that the PD effects themselves are fundamentally changed from off-resonance to on-resonance, beyond what is expected simply because the photoelectron kinetic energy changes. Our preliminary data exploring the variation of radiation polarization show very strong effects that at this point do not have a theoretical explanation; this is certainly worthy of further study.

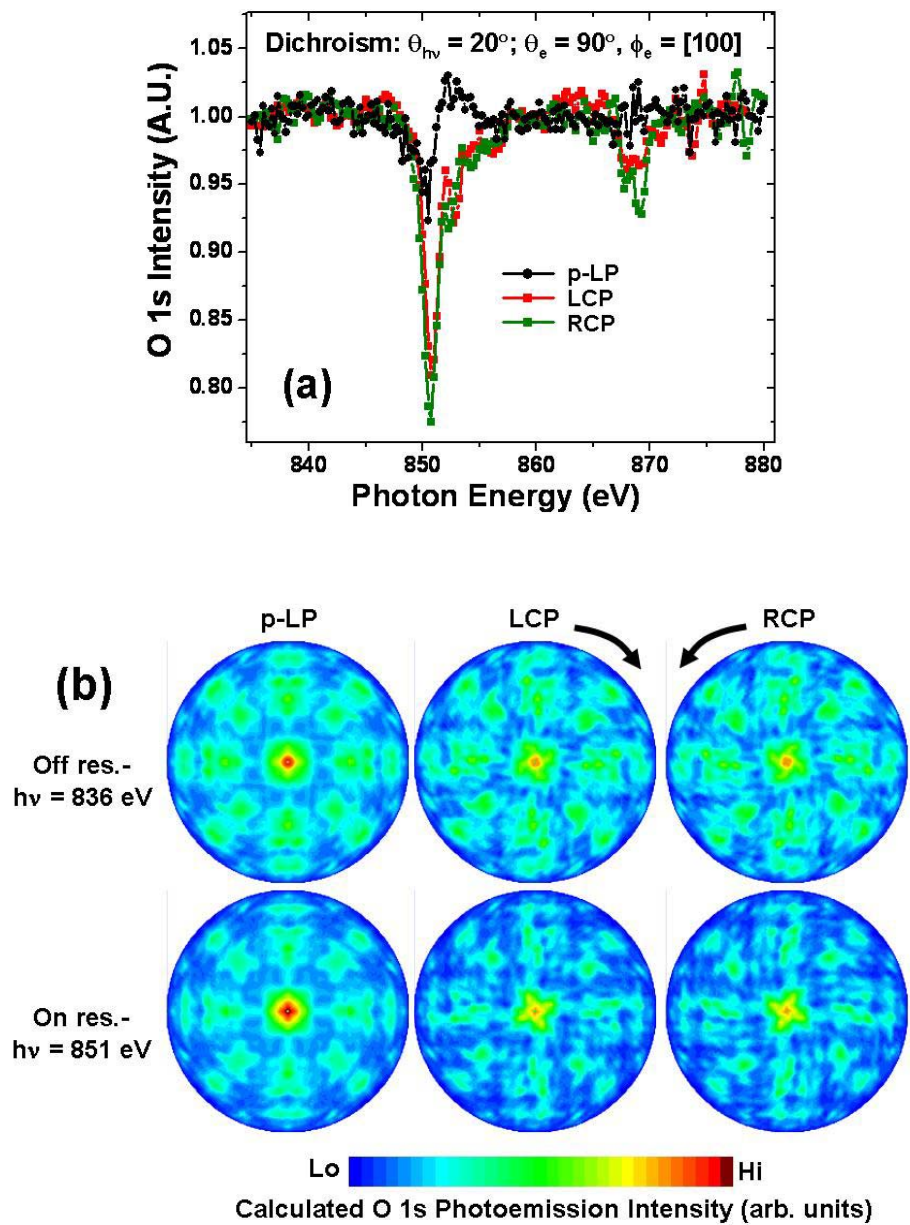

**Figure 5 (Color online):** (a) Measured effect of varying radiation polarization from p-type linear (p-LP) to right circular (RCP) and left circular (LCP) on the MARPE effects in O 1s emission from NiO, with a 20° incidence angle and normal photoelectron emission (i.e., for the same geometry as the data in Fig. 3(c)). (b) Comparison of calculated full-hemisphere photoelectron diffraction patterns from a cluster of 150 atoms for excitation with a 20° incidence angle and with p-LP, RCP, and LCP excitation. Two sets of patterns are shown, for off-resonance and on-resonance excitation energies

We have so far considered MARPE here to involve photoemission from one core level while passing through resonances of a core level on another atom, but analogous effects in valence-level emission on passing through a weakly-bound core level resonance from another atom have been both observed experimentally[31,32,33,34] and discussed theoretically.[35,36,37] These effects have been observed experimentally in measurements near solid-solid interfaces (in which the interatomic effect occurs across the interface),[31,32] in emission from a molecular orbital in a free molecule,[33] and in valence emission from clusters of atoms.[34] In the context of free molecules and atomic clusters, these effects have been termed interatomic coulomb decay (ICD), with a theoretical model having been elaborated[35] which is equivalent to that proposed previously for MARPE, but which considers also the interatomic Auger process, and for which the lower energies of excitation permit assuming that the wavelength of the exciting radiation is large with respect to the atomic distances involved. By contrast, a previously discussed theory of MARPE has used a more general fully-retarded description of the radiation field.[4,7] We thus expect that similar interatomic resonant photoemission effects will be found in many other systems, as for example, endofullerenes in which a certain atom is encapsulated in a carbon-based cage, and that such effects will constitute a new probe of electronic structure and near-neighbor atom identities in complex materials.

Finally, we note in closing that the x-ray optical model discussed above can be extended to describe fluorescent x-ray emission. For the case of a fluorescent energy that is far from any resonance and at a fluorescence exit angle $\theta^F$ that is large enough to minimize refraction and reflection at the surface, this involves simply replacing $\Lambda_e sin\theta_e$ with $\Lambda_x^F sin\theta^F$ in Eqs. (1), (2) and (4), with $\Lambda_x^F$ equal to the fluorescent x-ray attenuation length along path length or $\lambda_x^F/[4\pi\beta^F]$ in obvious notation. At this level of the theory, MARPE is dominated by what is usually termed the self-absorption effect in x-ray emission, with a more approximate, but standard, model for this effect having been used in two recent discussions of interatomic effects in x-ray emission[6,16]. However, this direct connection of MARPE and self-absorption was not realized in some earlier work[16].

**Acknowledgements:** This work was supported by the Director, Office of Science, Office of Basic Energy Sciences, Materials Science and Engineering Division, U.S. Department of Energy under Contract No. DE-AC03-76SF00098, and the IBM-Infineon technologies joint MRAM research project.